# Comet McNaught (260P/2012 K2): spin axis orientation and rotation period.

**Federico Manzini▼\*\*, Virginio Oldani▼\*\*, Roberto Crippa\*\*, José Borrero\*\*\*, Erik Bryssink[#], Martin Mobberley[##], Joel Nicolas[###], Sormano Obs.[x]**

[▼] SAS, Stazione Astronomica, Sozzago, Italy (IAU A12)
[\*\*] FOAM13, Osservatorio di Tradate, Italy (IAU B13)
[\*\*\*] Borrero Observatory, Michigan, USA
[#] BRIXIIS Observatory, Kruibeke, Belgium (IAU B96)
[##] Cockfield, Suffolk, UK (IAU 480)
[###] Les Mauruches Obs., Vallauris, France (IAU B51)
[x] Sormano Observatory, Italy (IAU 587)

e-mail: manzini.ff@aruba.it

## ABSTRACT

Extensive observations of comet 260P/McNaught were carried out between August 2012 and January 2013. The images obtained were used to analyze the comet's inner coma morphology at resolutions ranging from 250 to about 1000 km/pixel.
A deep investigation of the dust features in the inner coma allowed us to identify only a single main active source on the comet's nucleus, at an estimated latitude of -50° ± 15°.
A thorough analysis of the appearance and of the motion of the morphological structures, supported by graphic simulations of the geometrical conditions of the observations, allowed us to determine a pole orientation located within a circular spot of a 15°-radius centered at RA = 60°, Dec = 0°. The rotation of the nucleus seems to occur on a single axis and is not chaotic, furthermore no precession effects could be estimated from our measurements.
The comet's spin axis never reached the plane of the sky from October 2012 to January 2013; during this period it did not change its direction significantly (less than 30°), thus giving us the opportunity to observe mainly structures such as bow-shaped jets departing from the single active source located on the comet's nucleus.
Only during the months of August 2012 and January 2013 the polar axis was directed towards the Earth at an angle of about 45° from the plane of the sky; this made it possible to observe the development of faint structures like fragments of shells or spirals.
A possible rotation period of 0.340 ±0.01 days was estimated by means of differential photometric analysis.

## 1. INTRODUCTION.

High resolution images of the inner coma of a comet show different structures, resulting from the presence of active dust-emitting regions on the surface of a rotating nucleus (Sekanina, 1987). The study of the morphology of these structures, as well as of their variations over time, can provide important information to understand the evolution of the active areas on the nucleus and to identify its properties, such as the rotation and the direction of the axis (Jewitt, 1997; Chen et al., 1987; Yoshida et al., 1993; Kidger, 1996; Samarasinha et al., 1997).
In fact, assuming the case of a continuous outflow of dust occurring from an active source placed at mid-high latitude, the rotation of the nucleus will cause the formation of a cone in the space, whose surface is formed by the emitted particles. Since on images taken from the Earth we can only observe the two-dimensional projection of this cone, its edges will be seen brighter (due to the contribution of all the particles along the line of sight) and will thus appear as two jets located in symmetric position on the nucleus.
The direction of the spin axis may in this case be assumed to coincide with the bisector of the angle formed by two symmetrical jets and passing through the center of the nucleus. If the emission cone is seen from one side it will appear as a fan, with the bright edges curved depending on the trajectory of the particles, which is in turn based on their physical properties. The emission of dust particles will instead take the shape of spirals or shells if the spin axis is directed towards the point of view. In this case, a continuous outflow will create a spiral, while an activity modulated by the insolation will create concentric circle arcs, or fragments of spirals, moving sunwards.

The determination of the rotation period of a comet through the study of the geometrical properties of its nucleus is a more complex task. It can be estimated through a continuous monitoring of morphological changes in the jets or shells arising from active areas on the comet's nucleus, depending on their geometry and on the geometry of the apparition. In some cases, when the presence of very active jets is observed, differential photometry may be applied to long series of images taken over several consecutive nights, to determine cyclical variations that may thus be assumed to correspond to the rotation period.

The study of the evolution of jets and shells in the inner coma of comet Hale-Bopp allowed us and other Authors to discover a correlation with the rotation of the nucleus (Schwarz et al., 1997; Sekanina et al., 1999; Samarasinha, 2000; Manzini et al., 2001). Comet Ikeya-Zhang was also found to be particularly suitable for similar morphological studies (Manzini et al., 2006), and later an analysis of the morphological structures of the inner coma of comet Machholz once again provided precise indications about the direction of its rotation axis (Farnham et al., 2007a; Manzini et al., 2012) and about its rotation period.

Comet 260P (McNaught) was discovered by R.H. McNaught with the 0.5 m Uppsala Schmidt telescope at Siding Spring (Marsden, 2005a) in the morning of May 20, 2005. The comet was initially named C/2005 K3, and once its orbit was determined it became clear that it was a member of the Jupiter family, with an orbital period initially calculated in 4.25 years (Green, 2005).

A redefinition of its period of revolution as a result of subsequent astrometric observations, led to a new, more precise determination of the orbital parameters, with a corresponding period of 7.11 years (Marsden, 2005b), which would have brought comet McNaught back at perihelion in September 2012. The comet was in fact found before this passage by M. Masek on May 18, 2012 and was called C/2012 K2 and then 260P/McNaught (Green, 2012). Recently, a more precise determination of the period and of the orbital elements has been reported by Williams (Williams, 2012), referring to an arc of orbit with 520 astrometric positions; the period was calculated in 7.07 years, and the perihelion passage was set for September 12.52, 2012.

No authors attempted to determine the pole position and rotation period of comet 260P during its first passage in 2005. Therefore, the objectives of our work were:
1. to determine the orientation of the spin axis of the nucleus through characterization of the details of the inner coma;
2. to determine the comet's rotation period by means of differential photometry of the brightness of the false nucleus and of the inner coma, as well as through observation of the displacement of dust features.

## 2. METHODS

Despite the favorable geometric conditions of appearance for the northern hemisphere during its passage in 2012, the comet has never become brighter than 14th magnitude, probably due to a low emission and a small number of its active areas. The highest luminosity was reached in the second half of September, when the comet passed at the perigee at 0.5836 AU, in the days immediately after its perihelion passage; at that time it was almost in opposition to the Earth and this allowed extensive observation sessions. Our first observations of comet 260P date back to July 27, 2012, when it was as far as 0.92 AU from the Earth and 1.60 AU from the Sun, but already active and with a coma estimated to be at least 20 arcsecs wide, corresponding to about 19,000 km. The geometrical conditions of the observations remained favorable throughout the transit of the comet, which was rising in declination in the constellation Andromeda (very high at northern latitudes) and was almost always at the meridian in the middle of the night.

In total, the comet was observed for over 42 nights in the period from July 2012 to January 2013, and extensively studied on consecutive nights from 21 to 24 October 2012. Further long series of observations were made on the nights of 5, 7, 20 November and 3, 10 December 2012, in order to analyze also the photometric behavior of the comet's nucleus. All observations were carried out under good seeing conditions.

The images of comet 260P (McNaught) were taken by the observatories listed in **Table 1**. Most of them are located in Europe and the telescopes used have a spatial resolution (up to about 300 km/pixel) sufficient to investigate the morphology of the inner coma. Several studies were carried out successfully with instruments within this size range on comets Hyakutake (Schleicher et al., 1998; Eberhardy et al., 2000), Hale-Bopp (Schwarz et al., 1997), Ikeya-Zhang (Manzini et al., 2006), Machholz (Lin et al., 2007; Manzini et al., 2012).

To the purposes of this research, additional images were taken with larger instruments at the Faulkes North telescope (Haleakala, Hawaii) on November 11, 2012 and at the TNG (Telescopio Nazionale Galileo, Canary Islands) during the first decade of January, 2013, with a resolution of 150 and 230 km/pixel, respectively.

The list of all the images on which our research is based, by date and with full description of their characteristics, is shown in **Table 2**.

### 2.1 *Image Processing*

All our data consist in CCD images taken with R-broadband filters or without filters to favor an increase in the signal-to-noise ratio for dust, thus minimizing the contribution of the presence of gases (Farnham et al., 2007b). The images were first processed and calibrated by applying conventional methods of data reduction (bias, dark and flat-field images, taken on the same night of the observing sessions).

For the purpose of studying the coma morphology, whenever a long series of images was taken on the same night, we processed co-added images centered on the bright central peak at a sub-pixel level (the optocenter), to maximize the signal-to-noise ratio. The time lag between our co-added images never exceeded 30 minutes. We validated this procedure by verifying that all the details in the co-added images were the same as, and in the same position of, those in every single frame; therefore, no smearing effect was evident. A further validation of this procedure was done by dividing a first 30-min series of co-added images by a second series of co-added images taken in the following 30 minutes to ensure that the details highlighted in the inner coma were always stable.

Further processing was performed by means of the *Larson-Sekanina* spatial filter (Larson and Sekanina, 1984) and of the *radial gradient* technique, to highlight the presence of possible radial features (i.e. jets) and/or haloes (i.e. shells) in the inner coma. The results of these two image treatments were compared with each other to confirm the findings.

The *Larson-Sekanina* algorithm was applied taking the optocenter as the reference point, assumed to correspond to the comet's nucleus. A rotational shift of $\theta = \pm 45°$ was chosen as it allows to detect macro-details with the highest contrast with respect to the background, without introducing artifacts susceptible to be misinterpreted as real structures, contrary to what often happens with smaller angles especially for low surface brightness objects. No radial shift was applied, as it did not show any additional detail. In the images the black and white streaks are star trails, the black streaks being created by the digital image processing.

The *radial gradient* technique is based on the idea that any radial outflow (i.e. a jet) leads to slightly enhanced levels of local brightness. As these enhanced radiance levels only account for a small percentage of the mean intensity, they are almost invisible without additional processing. In order to detect these differences, we apply a *radial gradient* technique that divides each pixel of single, concentric, 1-pixel wide rings located around the center of brightness (i.e., the assumed nucleus position) by the median of all pixels in the ring. The process is repeated circularly over the whole image, for all subsequent rings. In the case of a perfect symmetry, no differences are noticed. On the contrary, if local jet structures result in locally increased intensity levels, these differences can be visualized after applying this technique. A similar technique was previously reported by Schleicher and Woodney (2003).

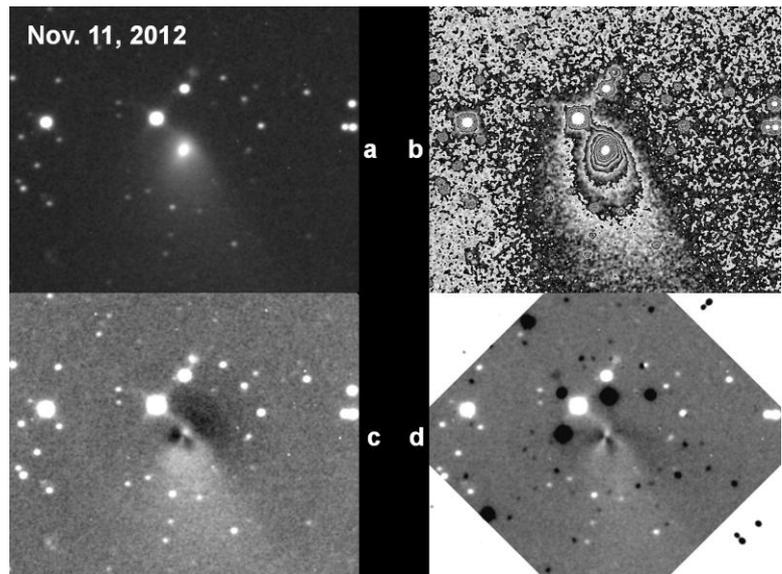

**Figure 1**
The original image (**a**) taken on Nov. 11, 2012, is shown to compare our image processing techniques: with superimposed isophotes (**b**), after applying a radial gradient (**c**) and after applying a $\theta = \pm 45°$ Larson-Sekanina algorithm (**d**). The images in (**c**) and (**d**) provided similar results, as well as in all other observing sessions during the apparition of comet 260P in 2012-13.

The comparison between the two above image processing techniques provided always identical results as shown in **Figure 1**, therefore all images presented in this paper are those treated with the Larson-Sekanina algorithm as they show the structures with the highest contrast.

We decided to analyze all the images in our possession, in order to create a sort of 'atlas' of the observed features, although we describe in this paper only those data that we consider significant to provide an overview of the apparition of comet 260P during its perihelion passage in 2012 (**Figure 2**).

## 2.2 *Analysis of the radial features and determination of the spin axis direction*

Since the details that are visible in the inner coma of a comet may look different at different resolutions and with different telescopes, all our images were first resized so that all had the same resolution of about 300 km/pixel.

Subsequently, each image was submitted to polar coordinates transformation centered on the central peak of brightness (i.e. the assumed position of the nucleus) in order to emphasize the radial details, with the θ angle on the horizontal axis (0° degrees corresponding to West, North at 90°), and the distance from the optocenter, in Km, on the vertical axis.

Photometry was then measured on three lines parallel to the horizontal axis at 5, 10 and 15

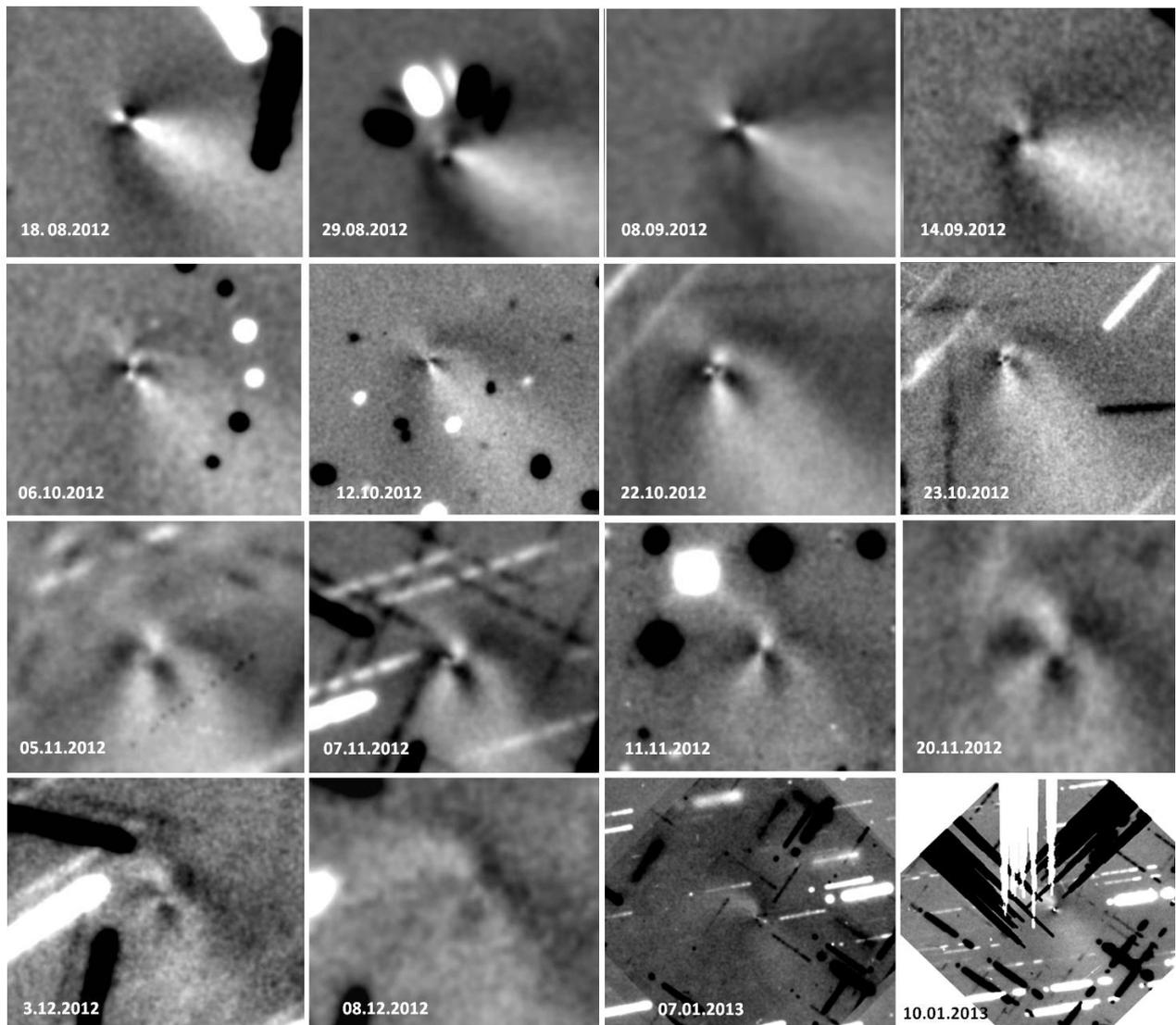

**Figure 2**
Comet 260P (McNaught) was followed throughout its appearance between August 2012 and January 2013.
The images of each observing session were co-added and processed to highlight the morphological details in the inner coma. The 16 panels show the evolution of the details. The comet passed at perihelion on September 12 and at perigee on October 4, 2012. Black and white streaks are due to the star trails enhanced by our treatments.

pixels (i.e. 1500, 3000 and 4500 km) on the distance axis, the highest peaks corresponding to the brightest radial features (tail and jets). This allowed precise calculation of the position angle of each jet, with an estimated error of less than 10 degrees.

Thereafter, all the bisectors of the angle formed by two symmetrical jets identified as above, passing through the center of the nucleus, were calculated, thus providing a good estimate of the direction of the spin axis on each image. An example of the results of the entire workflow is shown in **Figure 3**.

### 2.3 *Graphic simulations*

Graphic simulations of the geometrical conditions of the observations of the comet's nucleus (such as pole direction in RA and Dec) at different dates, were compared with the images taken on the same dates in order to verify that the morphologic features visible in the images were compatible with the physical parameters set for the nucleus. In these simulations the cometary nucleus was assumed to be spherical for simplicity, even though the comets visited by probes (1P/Halley, 19P/Borelly, 81P/Wild 2, 9P/Tempel 1) or observed with radar techniques (8P/Tuttle) proved to have irregularly shaped nuclei and a roughly triaxial structure.

The simulations of the rotational state of the comet's nucleus, consistent with the hypothesis based on the observations, are shown in the figures 5-6 and 8-9 (left side, not in scale) side by side with the corresponding processed CCD images (right side). In the simulations, the apparent size of the nucleus as seen from the Earth, the presumed direction of its spin axis and all the geometric Sun-Earth-comet conditions were preserved. The simulations also show the effect of the sun light on the nucleus.

All graphic simulations were obtained with the software Starry Night Pro Plus v. 5.8.4 (Simulation Curriculum Corp., Minnetonka (MN), USA).

### 2.4 *Animations*

Images obtained on October 8.9, 2012 made it possible to create an animation of the motion of comet 260P across the stars while transiting near galaxy UGC1514 in Andromeda.

Based on the determination of the direction of the spin axis, we produced an animated simulation that shows the geometric pattern of the passage of comet 260P. The apparent size of the nucleus, the direction of its spin axis and all the geometric Sun-Earth-comet conditions are preserved as in the non-animated images; the cometary nucleus is observed in geocentric position and thus its size varies with the distance from the Earth.

A second animation shows a simulation of how the motion in latitude of the insolation on the nucleus would appear if observed from a fixed point about 500 km away from the comet.

These two simulations cover the period from August 1, 2012 to March 1, 2013; the comet's nucleus is supposed to be spherical and with a rotation period as we found by photometric analysis of our data. Each second of the animation corresponds to a time-lapse of 8 days. These animations can be found at: *www.foam13.it/Risultati_scientifici/Ultime_novità_scientifiche/260P/260p.htm*.

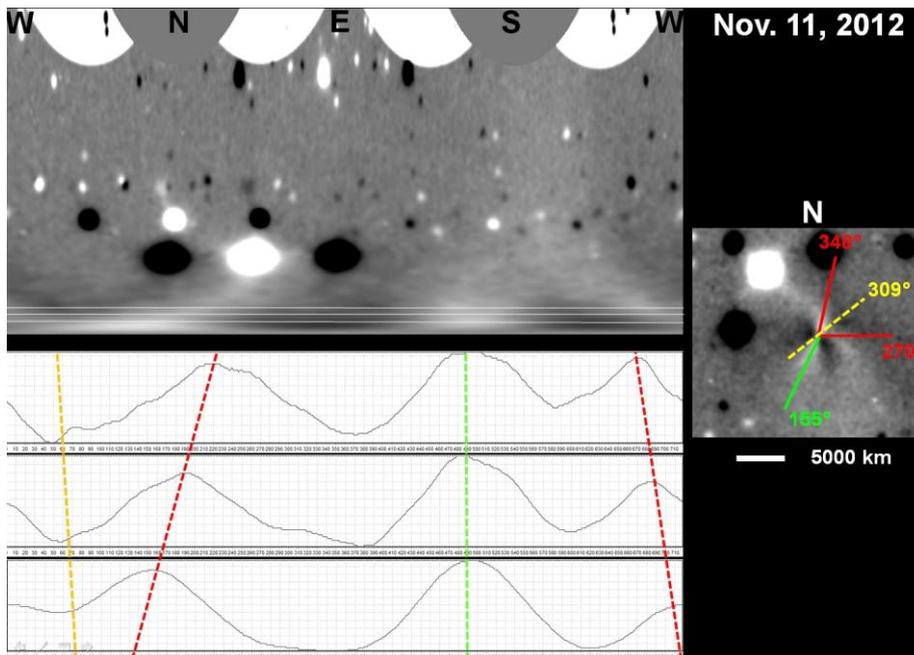

**Figure 3**
The images taken in each observation session were submitted to polar coordinates transformation centered on the central peak of brightness of the comet to emphasize the radial details, with the θ angle on the horizontal axis, and the distance from the optocenter, in pixels (corresponding to Km), on the vertical axis (top left). Photometry was done on parallel lines to the horizontal axis at 5, 10 and 15 pixels (i.e. 1500, 3000 and 4500 km, white lines on the image) on the distance axis to measure the precise position of the brightest radial features (dashed green line indicates the tail, red lines are the jets); all the bisectors of the angle formed by two symmetrical jets identified as above, passing through the center of the nucleus, were calculated to provide an estimate of the direction of the spin axis on each image (dashed yellow line). The tree photometric analyses on the bottom left side have a resolution of 2 pixels/degree.

### 3. SPIN AXIS ORIENTATION

Over the period between August 2012 and January 2013, by applying the image processing techniques described above, we could identify the presence of two symmetrical jets, most likely the result of the outflow of dust from a single active source on the rotating nucleus of the comet. From the analysis of the morphology of the inner coma, it was plausible to estimate the position of this source at mid latitude on the southern hemisphere of the nucleus (**$lat_{com.}$ = -50° ± 15°**).

The position angles of the spin axis were measured on each image taken over the observation period with the method described in section 2.2. The graphic representation of all the measured positions is shown in **Figure 4**.

Using a trial and error approach, by systematically varying the position of the spin axis over the sphere of the nucleus in the graphic simulations, we determined its orientation as the simulated projected position on the plane of the sky that best fitted with the measured angles. The best fit for the direction of the spin axis was estimated to be approximately at **RA = 60°** and **Dec = 0°**, falling within a circular spot with a radius of 10° to 15°.

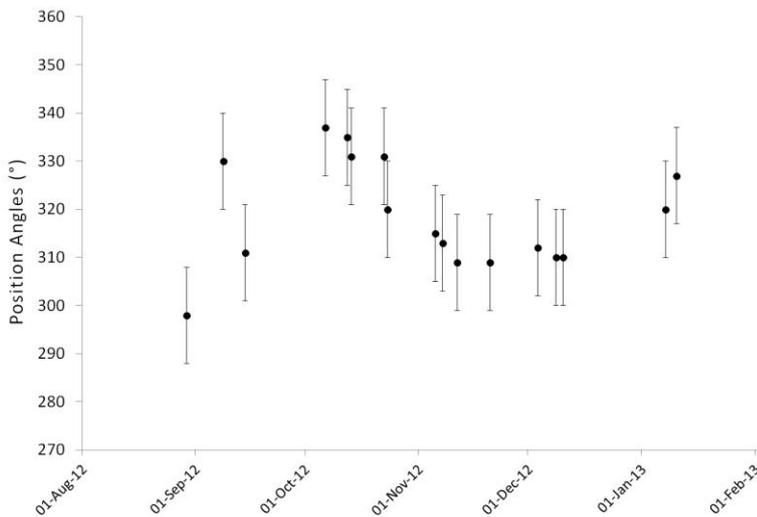

**Figure 4**
Plot of the position angles of the spin axis measured on each image taken over the observation period.

The images collected on September 14, October 12 (**Figure 5**) and 23, November 05, 11 (**Figure 6**) and 20, December 03, 08 and 10 resulted to be the most suitable to this analysis, as they were very representative of how the presence of two symmetrical jets should have appeared based on the geometrical position of the comet's nucleus with a direction of the spin axis only a little tilted with respect to the plane of the sky.

The simulations of the rotational state of the comet's nucleus were also compared with the images taken in August 2012 and in January 2013 (when we could take advantage of the large aperture of the TNG). These images do not show obvious symmetrical jests, but rather bow-shaped structures that we interpreted as most likely being fragments of a shell. Indeed, consistent with these observations and with what would be expected according to Sekanina's model (Sekanina, 1987), the graphic simulations did show the spin axis much more tilted with respect to the plane of the sky and oriented towards the point of view compared to how it looked like in the period from September to December 2012.

### 4. DESCRIPTION OF THE APPARITION OF COMET 260P

The comet was at perihelion on September 14, 2012 (R = 1.49 AU) and at the perigee on October 04, 2012 (delta = 0.58 AU). During the period from September to December the treatment applied to highlight the morphology of the inner coma showed a structure that is consistent over time with two opposing and nearly symmetrical jets, which appeared curved because of the action of the radiation pressure from the Sun and of the geometrical conditions of the observation. The angular size of the two jets suggests an origin from a single active area with poorly collimated emission.

The images obtained on November 11, 2012 (R = 1.63 AU, delta = 0.70 AU, Faulkes telescope at Haleakala, Hawaii, USA) are representative of the morphology observable throughout the period. Here the tail is directed toward PA 155° and the two jets are symmetric with respect to PA 320°; the overall structure of the comet resembles the Greek letter "Tau". No effects due to the rotation of the nucleus (such as clumps of material) are visible (**Figure 6**).

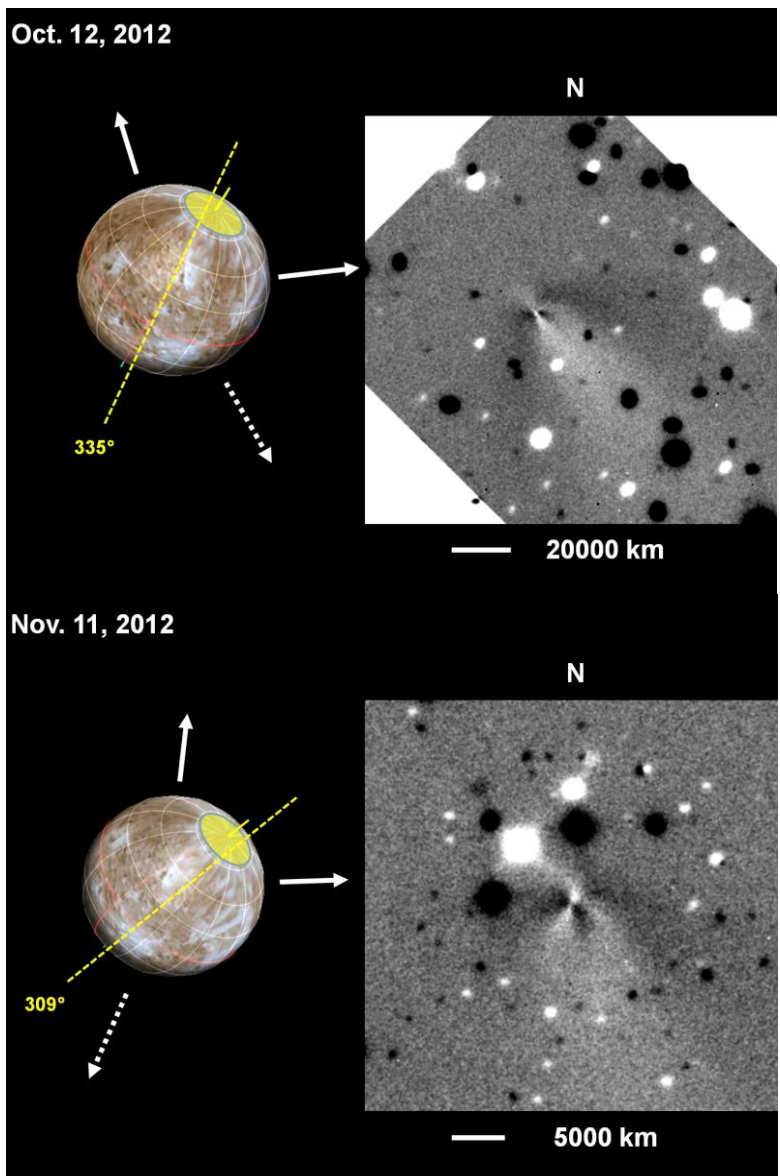

**Figure 5 and Figure 6**
The result after processing co-added images taken at two different dates (resp. Oct. 12 and Nov. 11, 2012) is shown on the right side of each figure. Black and white points are due to stars enhancement following image processing.
A simulation of the geometrical conditions of the observation is shown on the left side, with a direction of the spin axis towards RA=60° and Dec=0°. The comet appears as seen from the Earth. A grid of comet-centric coordinates and the rotation poles are also shown (yellow the south pole, blue the north pole); the size of the nucleus is scaled according to the distance from the Earth. The directions of the jets (solid arrows) are indicated as described in the paper. The tail outflow is shown as a dashed arrow; the direction of the Sun is opposed to this arrow. The superimposed dashed yellow line indicates the bisector of the photometrical measured positions of the two opposed jets; deliberately this line does not coincide with the axis of rotation (but it is parallel) in order not to conceal important parts of the simulation.
Centered on the rotation pole the error circle is shown (in pale yellow) where the true pole of the comet should fall. The comet appears to take the shape of the Greek letter *Tau*.

### 4.1 Emitting activity

The emitting activity of a comet should be modulated as a result of repeated cycles of "night and day" that occur at the latitude of an active source due to the rotation of the nucleus. Such a phenomenon can usually be observed in processed images in the form of clumps of material along the whole extent of the jets.

No apparent clumps are visible in our images, except for those taken on October 22 and 23, 2012 (R = 1.56 AU, delta = 0.61 AU). Here a clump of material is visible for the first time inside the jet heading westward. The observations, carried out in two consecutive days, should have shown a shift of the clump away from the comet's nucleus; the centroid is, however, very difficult to assess because the clump is quite diffuse and poorly condensed.

**Figure 7** shows the result after processing the original exposure taken on October 23, 2012 by means of a treatment with a median radial gradient associated with a 45° Larson-Sekanina filter. A 230-pixel long line superimposed on the two jets corresponds to the photometric analysis displayed in the underneath graph. The value of the pixels with respect to an arbitrary value of the sky background is shown on the ordinate. The brightness peak value is set to x = 115 and the clump (double?) shows an extension of about twenty pixels around x = 165.

### 4.2 Observation of morphological structures in August 2012

During the month of August, as the comet was approaching the Sun and the Earth, in addition to the tail, directed towards PA 250°, a curvilinear jet could be observed, followed by a (spiral) track that was developing from PA 90° progressively towards PA 180°, with westward concavity (**Figure 8**).

This feature became visible after the radial gradient and the rotational processing were applied; it could possibly represent the beginning of a spiral track or of a shell left by the outflow from an active source, with the comet's spin axis very divergent from the plane of the sky (**Figure 9**).

In this image the south pole of the comet's nucleus is facing the Earth; the appearance of the developing spiral is compatible with the estimated position of the spin axis of a nucleus rotating in a clockwise direction. Examples of such morphological structures were very evident in comet Hale-Bopp (Jewitt D., 1997, Manzini et al., 2001). Unfortunately, the presence of a bright star (white and black dots) just close to the comet prevented a further and more refined processing of this image.

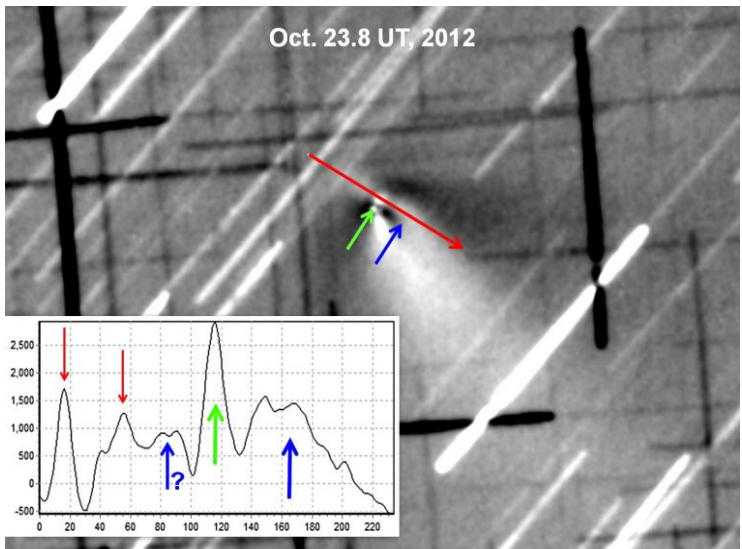

**Fig. 7**
The co-added images taken on October 23, 2012 are shown here after more intensive processing to better show the inner details in the coma. A line is drawn across the brightest features of the two symmetrical jets. A photometric analysis of the 230-pixel line is presented at the bottom: the jet to the right is brighter than the counterpart on the left and presents photometric irregularities due to the presence of a clump of material in position x = 170, about 20-pixel wide, which might consist of an outflow of material modulated by the rotation of the nucleus. Other bright peaks appearing in the photometric analysis (i.e. at position x = 15 and x = 55, indicated with a red arrow) are due to the presence of star traces in the field. Another blue arrow and a question mark indicate the possible counterpart of the clump, maybe present on the eastern jet. The position of this clump seems to be drifting apart from the nucleus in the following days; it is almost impossible to determine the extent of displacement because the material is very diffuse, without a marked condensation. The brightness scale is arbitrary.

### *4.3 Observation of morphological structures in January 2013*

Since the comet was moving far away, we took some images in two dates, January 07 and 10, with the TNG (Telescopio Nazionale Galileo at Canary Islands) to collect the latest data suitable to confirm all the simulations and the assumptions made with the previous images. The resolution obtained with this telescope (3.58 m in diameter) was almost 3 times better than that in the previous dates and the signal collection much higher. The image processing has been done in an identical manner to allow a consistent analysis.

Two small, roughly collimated jets were outflowing from the nucleus; a diffuse fan of material was expanding from them, evident in the treated images. The two structures were almost symmetrical with respect to PA 327°, that is approximately the direction of the spin axis, in good agreement with our simulation derived from the determined direction of the axis of rotation (**Figure 10**).

After an extreme stretching was applied to the original image, a second fan, weaker than the previous one, could be observed in the northern area of the nucleus and directed eastward. This latter fan could be made of material that had been ejected from the active source during a rotation of the nucleus previous to that which created the most intense fan. For this reason, it appeared farther from the nucleus and more widespread. In **Figure 11**, the graph of a photometric analysis, along the indicated line, has been superimposed; the two peaks of brightness due to the presence of the two fans are evident.

## 5. DETERMINATION OF THE COMET'S ROTATION PERIOD.

### *5.1 Analysis through photometry.*

Comet 260P/McNaught was extensively observed at SAS and FOAM13 observatories over several hours during the nights of October 22 and 23, November 05, 07 and 20, December 03 and 10 with the aim of determining any photometric variations of the nuclear region in the red filter (**Table 3**).

More than 3000 frames were analyzed by means of differential photometry with respect to the stars included in the field of view; we used the same stars on the same night also for the observations conducted simultaneously with two different telescopes, obtaining comparable light curves.

We analyzed all images with aperture photometry; the instrumental magnitudes were measured by means of a circular window with a 6-pixel radius (corresponding to 4.1 arcsecs) at SAS observatory and 3-pixel radius (corresponding to 4.5 arcsecs) at FOAM13 observatory, centered on the comet's nucleus (the optocenter), and by means of a concentric annulus with inner and outer radii of 60 and 70 pixels, respectively, to detect the median sky level. These sizes were chosen given that no seeing effects were visible in the photometry, as the measurement aperture was always larger than the FWHM of the stars in the nights of observation. The size of the central window allowed us to take into account any variations in the brightness due to the features located on the rotating nucleus (up to a distance between 2000 and 2500 km) and, at the same time, to new material emitted from the active source, and including little or no contribution from the coma. The size of the outer ring guaranteed to include only very faint parts of the tail, which was thus bringing no contribution to the photometric analysis.

Actually, in aperture photometry changes in the seeing conditions may affect the contribution of the coma, which in turn can dominate variations in the light curve even when the comet is unchanging. However, the stability and accuracy of the curves obtained during each observing session allowed to quantify the changes in magnitude due to seeing as minimal, since every image had an exposure time of 60 seconds, i.e. well beyond the time of instantaneous variation of the waves of atmospheric turbulence in our observing sites, and were therefore averaged over the course of the exposure.

This method seems to be efficient for the analysis of active comets presenting few emitting jets: this was the case for comet Machholz (Reyniers et al., 2009; Manzini et al., 2012), 8P and 46P (Manzini et al., 2013), 10P (Knight et al., 2011), 176P (Hsieh et al., 2011). The same method has been successfully applied for the photometric analysis of comets far from perihelion where, however, the bare nucleus could be seen without a coma around.

All the photometric data available for this analysis are shown in **Figure 12**, whilst in **Table 3** the geometric conditions during the photometric sessions are listed together with the colors that have been associated with the symbols on the figure.

The photometric data were subjected to a frequency analysis by means of the Bloomfield's method (Bloomfield, 1976). The Bloomfield method calculates a power spectrum, starting from unequally-spaced data, using the Least Squares Standard Technique.

The whole periodogram resulting from the analysis with the Bloomfield's method is shown in **Figure 13**; prominent periods in our Period Window appear as peaks.

The best period that could be derived from the highest peak at the frequency of rotation F = 2.9978 was P = 0.334 ± 0.004 days (8.06 ± 0.1 h)

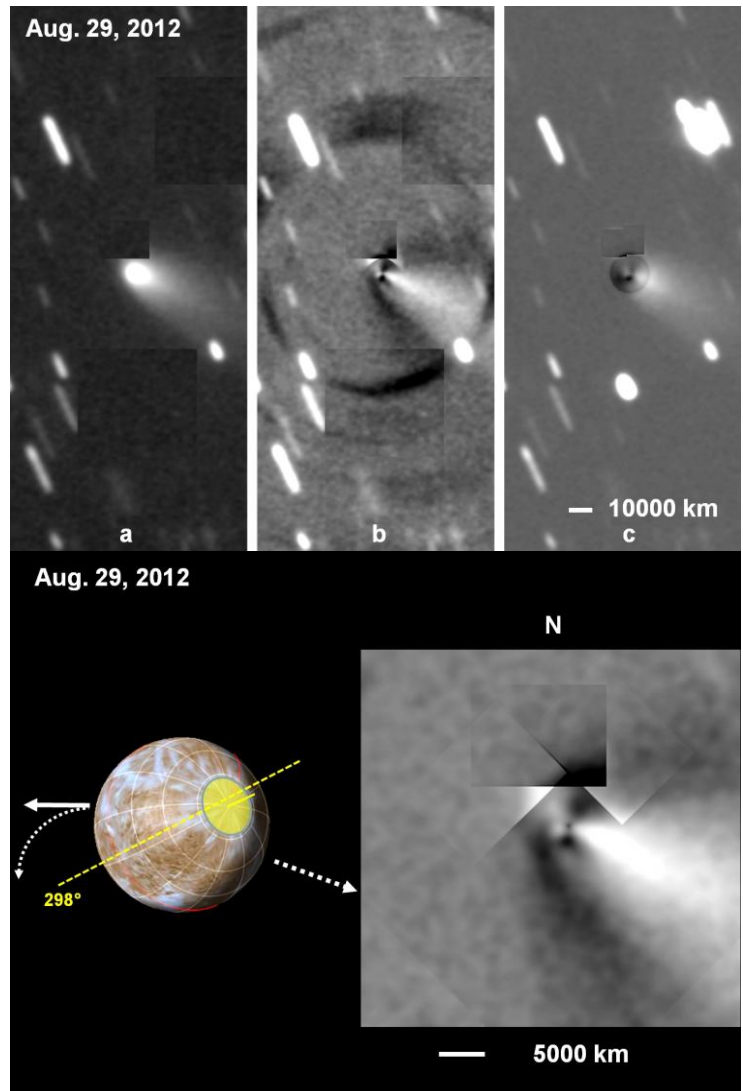

**Figure 8 and Figure 9**
In August the comet was approaching the Sun and the Earth.
According to our simulation, the direction of the rotation axis of the comet and its angle with the plane of the sky should have favored the visibility of curvilinear structures or shells, due to the outflow of dust from active sources on the nucleus. In fact, the processed images show signs of a curvilinear jet (or of a fragment of a shell) followed by a (incomplete spiral) track that was developing from PA 90° progressively towards PA 180°, with westward concavity. The structure of the morphology suggests that we are looking at the south pole of the comet. This feature became visible after the radial gradient and the rotational processing were applied (frame **a** = original co-added image, **b** = image treated with the Larson-Sekanina algorithm, **c** = image treated with the removal of the radial gradient). The tail is directed towards PA 250°.
The brightest stars in the original images have been covered with boxes.

on a time span of 49 days. The photometric measurements have been merged according to this period in the light curve shown in **Figure 14**.

Two additional periods: P = 0.341 days (8.18 h) and P = 0.350 days (8.4 h), were provided by the analysis with similar probability, therefore a clear-cut choice between them was not possible. However, considering their extreme proximity in time, a rotation period of the nucleus **P = 0.340 ± 0.01** days appears to be a reasonable estimate.

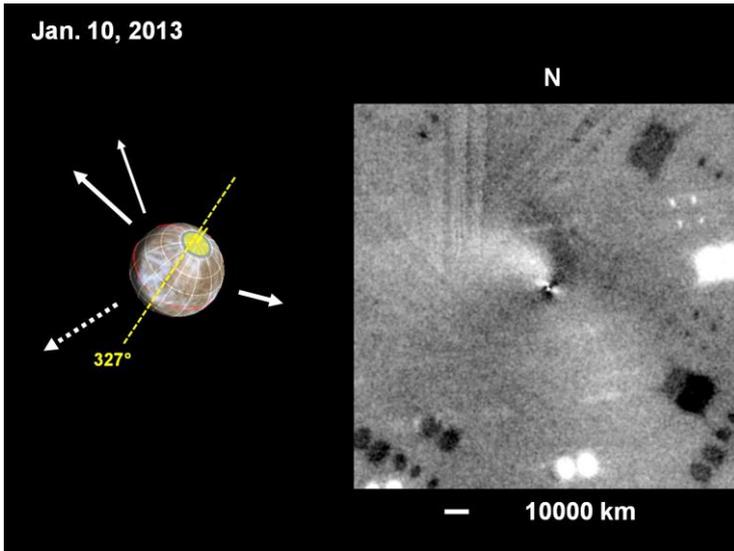

**Figure 10**
High resolution images taken on January 10, 2012 with the TNG telescope show that, compared to earlier dates, the brightest "jet" is now that placed to the east of the nucleus. Other morphological structures are also visible.

### 5.2 Analysis through observation of the displacement of dust features.

The observation and analysis of the motion of repetitive structures such as halos, shells or clumps of material outflowing from a comet's nucleus allow the determination of its rotation period; this method has been previously applied to estimate the rotation period of different comets, such as Hale-Bopp and Machholz (Samarasinha et al., 2011; Hsieh et al., 2011; Manzini et al., 2001, 2012; Waniak et al., 2012).

We could not determine the rotation period of comet McNaught through the analysis of the images taken over the study period. Clear morphological signs of a rotational activity were identified only on Jan. 7, 2013, when the comet was already quite far from the Earth and the Sun, and we could take advantage of the TNG capabilities. The images taken with the TNG on January 10th show the presence of two arcs of shell that were moving away from the cometary nucleus to the east (see also paragraph 4.3). These structures most probably developed from an outflow of material occurring over two different rotations of the comet nucleus, the farther (and more diffuse) corresponding to the emission of an earlier rotation. In images processed with a particularly intense stretching, two morphological structures associated to the outflow of material can also be observed to the west of the nucleus, but their distance from the nucleus is not measurable; they are probably due to an emission that occurred over the "half-rotation" of the nucleus (Period/2) previous to the one that created the structures on the east. An accurate measurement of the distance of the two arcs of shell to the east of the nucleus is difficult because there is no clear point of condensation inside them and they appear very diffuse; photometric measurements, repeated at multiple points, led to a value of about 11-13 pixels. The resolution of the TNG was 340 km/pixel on that date.

Based on a rotation period of 0.340d (±0.01d) = 29376s (±864s), an estimate of the minimum speed of the emissions from the active area on the nucleus, applying the formula:

$$V_{emission} \text{ [km/s] on the sky} = \text{Distance} / 29376$$

leads to a value of the expansion speed projected on the plane of the sky approximately between 0.13 and 0.15 km/s.

The expansion speed can be de-projected by applying a correction to the corresponding phase angle at the time of observation (**Table 2**):

$$V_{emission} \text{ [km/s] deprojected} = V_{emission} \text{ [km/s] on the sky} / \sin(\text{phase angle})$$

The application of this formula leads to a value of the expansion speed between 0.29 and 0.31 km/s.

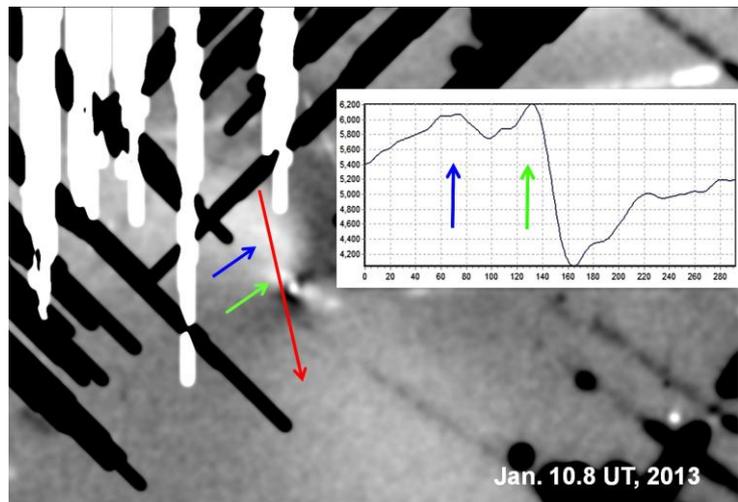

**Figure 11**
The photometric analysis of the 290-pixel long red line that crosses the coma (top to bottom, east of the nucleus) has been put into a chart. Two peaks of brightness at x = 70 and x = 130 are evident; they may indicate material emitted by the single active source over two rotations of the nucleus. The fan placed at x = 130 is closer to the nucleus and thus denser and brighter, while the second at x = 70 had more time to move away and is therefore fainter and more widespread. The brightness scale is arbitrary. The white and black strikes are due to CCD blooming coming from of a saturated star in the field of view.

Considering the geometrical conditions of the observations on those days, the resulting value is perfectly compatible with that found for the outflow of particulate matter on many other comets.

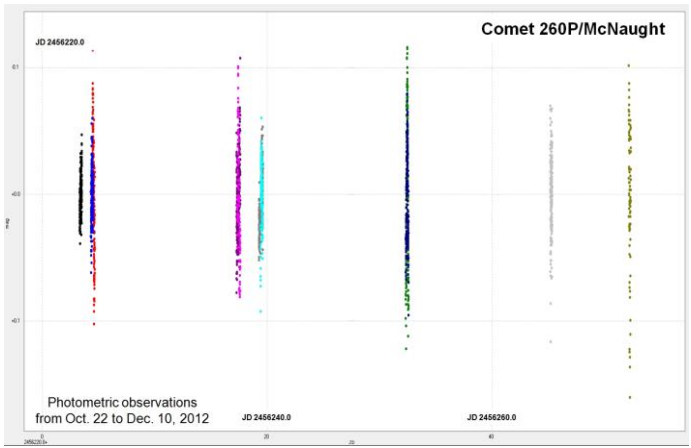

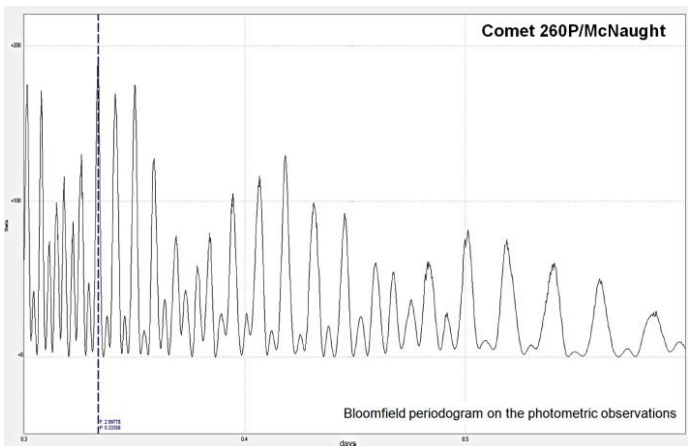

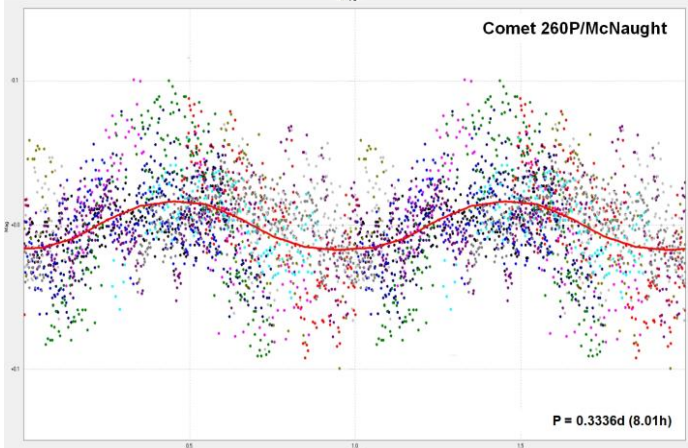

**Figure 12**
Photometric measurements on R-filtered images taken from October 22 to December 10, 2012.
**Figure 13**
Analysis of all photometric measurements of Figure 12 with the Bloomfield's method in time periods from 0.3 to 0.6 day, resulted in this periodogram, with the highest peak at frequency F = 2.9978/d.
**Figure 14**
The light curve corresponding to the period found with the Bloomfield periodogram (Figure 13) is quite compact, but the variation between the minimum and maximum brightness is of only 0.07 magnitudes. The period of this curve (P=0.334d) represents the rotation period of the nucleus.

## 6. FORECAST FOR THE RETURN OF COMET 260P IN 2019.

The determination of the direction of the spin axis of comet 260P/McNaught allows making some predictions for its next perihelion passage, which will happen in 2019 at 1.416 AU from the Sun; the comet will pass at perigee around the beginning of October 2019 with a minimum distance of 0.562 AU from the Earth.

The observation conditions will be favorable, with the comet very high on the horizon for observers located in the northern hemisphere, moving from the constellation of Triangulum to that of Andromeda. The comet will be in opposition around November 20, 2019.

The geometric conditions of the passage will be similar to those of 2012, and the comet will show to the Earth the same hemisphere already observed. On July 10, 2019 the comet will be at 1.00 AU from the Earth and 1.57 AU from the Sun, with a spin axis diverging about 10° from the plane of the sky: at that time the active area observed in the return of 2012 (at mid-low latitudes) should already produce material, and morphological configurations should already be observed in the coma (**Figure 15a**).

In contrast to the passage of 2012, the rotation axis of comet 260P will be roughly directed towards PA 270°, therefore it will have to be taken into account that the possible morphologies of the inner coma will vary their position angle accordingly.

The apparent direction of the spin axis will remain nearly the same from September until the end of 2019 (comet at perihelion on September 11, 2019); the lighting conditions on the nucleus will change slightly in favor of a greater insolation in the southern hemisphere (**Figure 15b**).

Between the end of January and February 2020 the spin axis will diverge from the plane of the sky and this would possibly allow the appearance of shells developing from the outflow of material (perhaps from the sole active source that has already been observed in the southern hemisphere during the passage of 2012). On January 10, the comet will be however as far as 1.92 AU from the Sun and 1.20 AU from the Earth (**Figure 15c**).

## 7. DISCUSSION

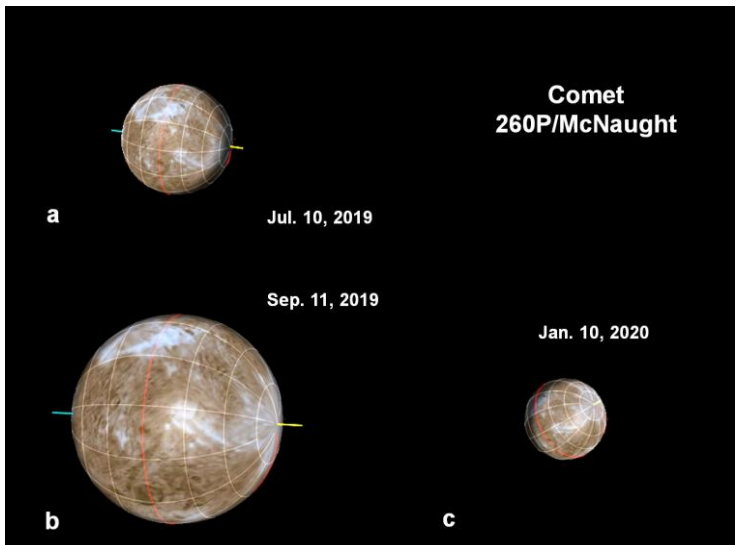

**Figure 15**
The determination of the orientation of the spin axis of the nucleus of comet 260P/McNaught allows simulating the geometric conditions of observation from the Earth for the next perihelion passage, which will take place in 2019. The nucleus will be greatly illuminated, and it will be possible to study any morphological structures of the inner coma related to the rotation of the nucleus. The apparent size of the nucleus is in agreement with the distance of the comet from Earth.

The analysis of the morphology of the inner coma of comet 260P/McNaught allowed us to determine some properties of its nucleus, such as the likely location of one major active area, the approximate direction of the spin axis and, presumably, the rotation period.

To date, no other authors have published analyses on comet 260P/McNaught during its two recognized perihelion passages in 2005 and 2012; therefore the results presented here cannot be compared with others.

The jets and features in the dust coma observed during the period of our observations could be explained by the presence of a single major active source on the surface of the nucleus. Simulations of the geometric conditions of the observation provided an explanation for the features observed in the inner coma: based on the average angle aperture of the symmetrical jets the active site appeared to be located at mid or low latitude ($lat_{com.} = -50° ± 15°$).

Fitting the images with our model allowed us to derive a direction of the spin axis of the nucleus located within a 15°-radius circular spot centered at RA = 60° and Dec = 0°.

This determination of the direction of the rotation axis resulted in an angle of about -35° ± 15° with respect to the comet's orbital plane and a longitude of 235° ± 15° at perihelion (0° in longitude is the Sun-comet vector). The rotation of the nucleus of comet 260P seems to occur on a single axis and is not chaotic, furthermore no precession effects could be estimated from our measurements.

We also obtained a possible value for the rotation period by means of differential photometry, by analyzing with the mathematical Bloomfield's method the magnitudes of the coma regions closest to the nucleus:

P = 0.340 ± 0.01 days, after prolonged observing sessions, time span 49 days.

However, the periodogram provided 3 possible solutions for the rotation period with similar probability: 0.334, 0.341 and 0,350 days; all these findings were within the limits of the error of the estimate of the photometric period. Although these results are all possible with similar confidence, in any case the rotation period appeared to be no greater than P = 0.35 days. The observed difference between the minimum and maximum light was of 0.07 magnitudes.

No results about the rotation period could be estimated by observing the wandering of dust features from the nucleus.

The future perihelion passage in 2019 will still be favorable for the observation of morphological structures in the inner coma, similar to those observed in the recent passage, if the emitting source identified in 2012 will turn on again.

## ACKNOWLEDGEMENTS

We would like to thank:
Emilio Molinari, director of TNG, Gloria Andreuzzi and the staff of Telescopio Nazionale Galileo for the great help provided in shooting the images with the TNG.
Patrizia Caraveo and Giovanni Bignami for the useful discussions.
Danilo Pivato for making his images of comet 260P available to us.
This research has made use of the LCOGT public Archive, which is operated by the California Institute of Technology, under contract with the Las Cumbres Observatory.

**Observations of Comet 260P/McNaught: Observatories and CCD**

| Code | Observatory | Lat. (°) | Long. (°) | Country | Telescope |
|------|-------------|----------|-----------|---------|-----------|
| IAU A12 | Stazione Astronomica Sozzago (SAS) | | | Italy | Cassegrain 0.4m, f/6,8 |
| IAU B13 | FOAM13 Obs. | | | Italy | Reflector 0.65m, f/5 |
| IAU B96 | BRIXIIS Obs. | | | Belgium | Reflector 0.4 m, f/3.7 |
| IAU 480 | Cockfield Obs. | | | United Kingdom | SCT 0.35m, f/7.7 |
| IAU B51 | Vallauris Obs. | | | France | Reflector 0.41m, f/3,3 |
| IAU 587 | Sormano Obs. | | | Italy | RCOS 0.5m, f/7 |
| Borrero | Borrero Obs. | 42°53'24"N | 85°35'24"W | Michigan, USA | SCT 0.2m, f/5.5 |
| IAU F65 | Faulkes N, LCGOT | | | Hawaii, USA | Reflector 2m |
| TNG | Telescopio Nazionale Galileo | 28°45'14"N | 17°53'21"W | La Palma, Spain | Reflector 3.58m |

| Code | CCD camera | Sensor | Pixel size (microns) | Resolution (arsec/pixel) | Filters |
|------|-----------|--------|----------------------|--------------------------|---------|
| IAU A12 | Hi-SIS43 | KAF-1603ME | 9x9 | 0.69 | 200 nm broadband, center at 650 nm |
| IAU B13 | Apogee Alta 1001 | KAF-1001ME | 24x24 | 1.52 | 200 nm broadband, center at 650 nm |
| IAU B96 | SBIG STL-6303E | KAF-6303E | 9x9 | 1.26 | broadband |
| IAU 480 | SBIG ST-9XE | KAF-0261 | 20x20 | 1.53 | broadband |
| IAU B51 | SBIG ST8XME | KAF-1603ME | 9x9 | 1.39 | broadband |
| IAU 587 | SBIG ST-11000 | KAI-11000 | 9x9 | 0.53 | broadband |
| Borrero | Artemis HSC | ICX-285 | 6.45x6.45 | 1.20 | broadband |
| IAU F65 | | FI CCD486 | 15x15 | 0.30 | Bessell R' |
| TNG | | E2V4240 CCD | 13.5x13.5 | 0.25 | SDSS G, SDSS R |

**Table 1:** Observations of comet 260P/McNaught: observatories and CCD cameras.

**Images of Comet 260P/McNaught**

| Date | Observ. | Mean time UT | N of images during obs. sessions | Tot. Exp. (s) | Mean FWHM of stars on images (pixel) | Scale (arcsec/pix) | Scale (km/pix) | Δ (AU) | R (AU) | Elong. (°) | Phase angle (°) | PsAng Vector (°) |
|---|---|---|---|---|---|---|---|---|---|---|---|---|
| **2012-07-27** | **IAU B51** | 01.30 | 6 | 1500 | 3.1 | 1.39 | 927 | 0.920 | 1.580 | 111 | 37.1 | 249.9 |
| **2012-08-18** | **IAU B51** | 00.10 | 1 | 300 | 3.1 | 1.39 | 751 | 0.745 | 1.522 | 119 | 35.4 | 249.2 |
| **2012-08-18** | **IAU B96** | 02.30 | 9 | 1620 | 3.5 | 1.26 | 680 | 0.745 | 1.522 | 119 | 35.4 | 249.2 |
| **2012-08-19** | **IAU B51** | 01.50 | 3 | 900 | 2.9 | 1.39 | 746 | 0.740 | 1.519 | 120 | 35.3 | 249.1 |
| **2012-08-19** | **IAU B96** | 03.28 | 5 | 900 | 3.5 | 1.26 | 676 | 0.740 | 1.519 | 120 | 35.3 | 249.1 |
| **2012-08-20** | **IAU B51** | 23.50 | 2 | 600 | 2.9 | 1.39 | 732 | 0.727 | 1.517 | 120 | 35.0 | 248.9 |
| **2012-08-24** | **IAU B51** | 00.50 | 2 | 600 | 2.8 | 1.39 | 708 | 0.703 | 1.511 | 122 | 34.4 | 248.2 |
| **2012-08-27** | **IAU B51** | 02.30 | 1 | 300 | 3 | 1.39 | 692 | 0.687 | 1.507 | 123 | 33.7 | 247.5 |
| **2012-08-28** | **IAU B51** | 02.30 | 3 | 900 | 2.9 | 1.39 | 686 | 0.681 | 1.505 | 125 | 33.5 | 247.2 |
| **2012-08-29** | **IAU B51** | 02.30 | 1 | 450 | 2.8 | 1.39 | 681 | 0.676 | 1.504 | 125 | 33.5 | 246.9 |
| **2012-09-08** | **IAU B96** | 02.00 | 7 | 1260 | 3 | 1.26 | 576 | 0.631 | 1.498 | 131 | 30.8 | 242.7 |
| **2012-09-14** | **IAU 587** | 20.49 | 2 | 600 | 5.2 | 0.53 | 235 | 0.611 | 1.497 | 134 | 28.9 | 238.9 |
| **2012-09-21** | **IAU B51** | 22.40 | 5 | 750 | 2.8 | 1.39 | 599 | 0.595 | 1.500 | 138 | 26.6 | 232.8 |
| **2012-10-02** | **IAU B51** | 22.50 | 2 | 600 | 2.9 | 1.39 | 587 | 0.583 | 1.513 | 144 | 22.9 | 219.1 |
| **2012-10-06** | **IAU 480** | 20.25 | 1 | 120 | 1.5 | 1.36 | 576 | 0.584 | 1.520 | 146 | 21.3 | 212.6 |
| **2012-10-07** | **IAU B51** | 23.25 | 5 | 1350 | 3 | 1.39 | 589 | 0.585 | 1.522 | 146 | 21.2 | 210.8 |
| **2012-10-09** | **IAU B51** | 23.30 | 5 | 1350 | 2.8 | 1.39 | 590 | 0.586 | 1.526 | 146 | 21.0 | 207.2 |
| **2012-10-10** | **IAU 480** | 19.12 | 1 | 120 | 1.7 | 1.41 | 601 | 0.588 | 1.528 | 148 | 20.4 | 205.3 |
| **2012-10-12** | **IAU 480** | 19.12 | 1 | 120 | 2.5 | 1.41 | 603 | 0.590 | 1.532 | 148 | 19.6 | 201.3 |
| **2012-10-12** | **IAU 587** | 20.40 | 1 | 300 | 3.5 | 0.53 | 227 | 0.590 | 1.532 | 148 | 19.6 | 201.3 |
| **2012-10-13** | **IAU B51** | 23.15 | 2 | 450 | 2.7 | 1.39 | 596 | 0.592 | 1.535 | 149 | 19.5 | 199.3 |
| **2012-10-13** | **Borrero** | 04.00 | 5 | 700 | 4.1 | 1.36 | 584 | 0.592 | 1.535 | 149 | 19.5 | 199.3 |
| **2012-10-15** | **IAU B51** | 23.50 | 5 | 1125 | 2.6 | 1.39 | 599 | 0.595 | 1.539 | 149 | 19.5 | 195.1 |
| **2012-10-21** | **IAU A12** | 19.50 | 25 | 1500 | 4.3 | 0.69 | 305 | 0.610 | 1.555 | 152 | 18.1 | 181.5 |
| **2012-10-22** | **IAU A12** | 23.20 | 212 | 11880 | 3.8 | 0.69 | 307 | 0.613 | 1.558 | 152 | 18.1 | 179.2 |
| **2012-10-23** | **IAU A12** | 23.10 | 210 | 8100 | 4 | 0.69 | 308 | 0.616 | 1.562 | 151 | 18.1 | 176.8 |
| **2012-10-23** | **IAU B13** | 23.50 | 400 | 15000 | 2.2 | 1.52 | 679 | 0.616 | 1.562 | 151 | 18.1 | 176.8 |
| **2012-10-24** | **IAU B51** | 20.30 | 4 | 720 | 2.7 | 1.39 | 624 | 0.619 | 1.565 | 151 | 18.1 | 174.5 |
| 2012-11-05 | IAU A12 | 21.15 | 130 | 9400 | 5 | 0.69 | 336 | 0.671 | 1.606 | 149 | 18.3 | 146.7 |
| 2012-11-05 | IAU B13 | 01.12 | 400 | 15000 | 2.3 | 1.52 | 739 | 0.671 | 1.606 | 149 | 18.3 | 146.7 |
| 2012-11-05 | IAU B51 | 22.15 | 3 | 675 | 3 | 1.39 | 676 | 0.671 | 1.606 | 149 | 18.3 | 146.7 |
| 2012-11-07 | IAU A12 | 23.35 | 160 | 9300 | 5 | 0.69 | 341 | 0.682 | 1.614 | 149 | 18.7 | 142.5 |
| 2012-11-07 | IAU B13 | 02.00 | 500 | 15000 | 2.2 | 1.52 | 751 | 0.682 | 1.614 | 149 | 18.7 | 142.5 |
| 2012-11-11 | IAU F65 | 08.40 | 5 | 300 | 7 | 0.3 | 154 | 0.706 | 1.630 | 147 | 18.9 | 134.5 |
| 2012-11-20 | IAU A12 | 23.30 | 240 | 10100 | 4.8 | 0.69 | 384 | 0.767 | 1.669 | 143 | 20.5 | 119.1 |
| 2012-11-20 | IAU B13 | 07.12 | 300 | 15000 | 2.5 | 1.39 | 773 | 0.767 | 1.669 | 143 | 20.5 | 119.1 |
| 2012-11-20 | IAU B51 | 20.00 | 5 | 1560 | 2.8 | 1.39 | 773 | 0.767 | 1.669 | 143 | 20.5 | 119.1 |
| 2012-12-03 | IAU A12 | 20.40 | 180 | 8000 | 5.4 | 0.69 | 439 | 0.877 | 1.732 | 137 | 22.9 | 102.9 |
| 2012-12-08 | IAU A12 | 20.00 | 90 | 5900 | 5.1 | 0.69 | 463 | 0.925 | 1.758 | 137 | 24 | 98.2 |
| 2012-12-08 | IAU B51 | 20.00 | 5 | 1140 | 2.8 | 1.39 | 932 | 0.925 | 1.758 | 137 | 24 | 98.2 |
| 2012-12-10 | IAU A12 | 21.40 | 100 | 6000 | 5.3 | 0.69 | 473 | 0.945 | 1.768 | 133 | 24.1 | 96.5 |
| 2012-12-28 | IAU A12 | 21.40 | 70 | 3000 | 2.7 | 1.38 | 1151 | 1.151 | 1.869 | 123 | 26.7 | 85.6 |
| 2012-12-29 | IAU A12 | 20.00 | 75 | 8000 | 2.7 | 1.38 | 1164 | 1.164 | 1.875 | 122 | 26.7 | 85.1 |
| 2012-12-30 | IAU A12 | 19.00 | 90 | 5600 | 2.6 | 1.38 | 1177 | 1.177 | 1.881 | 121 | 26.7 | 84.7 |
| 2013-01-07 | TNG | 11.00 | 10 | 1800 | 4.5 | 0.25 | 232 | 1.282 | 1.929 | 116 | 27.2 | 81.9 |
| 2013-01-10 | TNG | 12.00 | 10 | 1800 | 4.4 | 0.25 | 240 | 1.324 | 1.947 | 114 | 27.4 | 81.1 |
| 2013-01-11 | IAU B51 | 21.00 | 5 | 1045 | 2.9 | 1.39 | 1348 | 1.338 | 1.954 | 114 | 27.4 | 80.8 |

**Table 2:**
Observations data of comet 260P/McNaught.
Dates of observations and international codes of observatories are in column 2 and 3.
The mean FWHM measured in pixels for the stars on the images during each observing session are shown in column 6. An evaluation of the average atmospheric seeing during the night may be obtained by multiplying this data for the corresponding value in column 7.
The original scale of the observation, measured as resolution in km/pixel at the comet distance, is given in Column 8.
The SOT (Sun-Observer-Target) angle, the Phase Angle (the angle of illumination of the Target: 0° = fully lit, 180° = no lighting) as viewed from Earth, and the angle of the Vector Radii (connecting the Sun and the Comet) as viewed from Earth, are listed in columns 11, 12 and 13, respectively.
The measures in columns 9 to 13 have been calculated with the HORIZONS Web-Interface
(http://ssd.jpl.nasa.gov/).

**R-filtered images of Comet 260P/McNaught**
Geometric conditions during photometric sessions

| Date UT | N° of images during the night | D (AU) | R (AU) | l (°) | b (°) | Phase (°) | Colours on figures |
|---|---|---|---|---|---|---|---|
| 2012-10-22.82 to 23.05 | 212 | 0.6312 | 1.559 | 33.2 | 10.5 | 18.1 | black points |
| 2012-10-23.82 to 24.05 | 210 | 0.6163 | 1.562 | 33.8 | 10.7 | 18.1 | red points |
| 2012-10-23.97 to 24.20 | 400 | 0.6163 | 1.562 | 33.8 | 10.7 | 18.1 | blue points |
| 2012-11-05.87 to 06.17 | 134 | 0.6717 | 1.606 | 41.8 | 12.1 | 18.3 | violet points |
| 2012-11-05.77 to 06.12 | 452 | 0.6717 | 1.606 | 41.8 | 12.1 | 18.3 | dark violet points |
| 2012-11-07.80 to 08.13 | 160 | 0.6826 | 1.614 | 42.9 | 12.3 | 18.5 | grey points |
| 2012-11-07.78 to 08.14 | 500 | 0.6826 | 1.614 | 42.9 | 12.3 | 18.5 | cyan points |
| 2012-11-20.86 to 21.09 | 240 | 0.7677 | 1.669 | 50.4 | 13.5 | 20.6 | dark blue points |
| 2012-11-20.88 to 21.09 | 300 | 0.7677 | 1.669 | 50.4 | 13.5 | 20.6 | green points |
| 2012-12-03.88 to 04.07 | 180 | 0.8782 | 1.733 | 57.4 | 14.4 | 22.9 | grey points |
| 2012-12-10.71 to 10.98 | 105 | 0.9462 | 1.769 | 60.9 | 14.7 | 24.2 | light green points |

**D** = Comet-Earth distance
**R** = Sun-Comet distance
**l** and **b** = heliocentric ecliptical longitude and latitude of the comet
**Phase** = Sun-Comet-Earth angle

**Table 3:**
Geometric conditions during the photometric sessions on comet 260P/McNaught and legend of symbols in figures 12 and 14.